\documentclass[conference]{IEEEtran}
\IEEEoverridecommandlockouts
\usepackage{cite}
\usepackage{amsmath,amssymb,amsfonts}
\usepackage{algorithmic}
\usepackage{graphicx}
\usepackage{textcomp}
\usepackage{xcolor}

\usepackage[hidelinks]{hyperref}
\usepackage{booktabs}
\usepackage{multirow}
\usepackage{tabularx}
\usepackage{subcaption}
\usepackage{caption}
\usepackage{threeparttable}
\usepackage{xspace}
\usepackage{bm}
\usepackage[table]{xcolor}
\usepackage{mathtools}
\usepackage{svg}
\usepackage{todonotes}
\usepackage{pifont}
\usepackage{enumitem}
\usepackage{url}
\usepackage{array}
\usepackage[ruled,vlined,linesnumbered]{algorithm2e}
\usepackage{amsmath}
\usepackage{amsthm}
\theoremstyle{definition}

\newtheorem{lemma}{Lemma}

\newtheorem{example}{Example}
\newtheorem{definition}{Definition}

\newcommand{\model}{UniCom\xspace}

\def\BibTeX{{\rm B\kern-.05em{\sc i\kern-.025em b}\kern-.08em
    T\kern-.1667em\lower.7ex\hbox{E}\kern-.125emX}}
\begin{document}

% \title{Conference Paper Title*\\
% {\footnotesize \textsuperscript{*}Note: Sub-titles are not captured in Xplore and
% should not be used}
% \thanks{Identify applicable funding agency here. If none, delete this.}
% }

\title{UniCom: A Unified Framework for Community Search and Detection via Transfer Learning}

% \author{Anonymous}

\author{
\IEEEauthorblockN{
Yifan Zhu\textsuperscript{1},
Hanchen Wang\textsuperscript{2},
Wenjie Zhang\textsuperscript{1},
Alexander Zhou\textsuperscript{3},
Ying Zhang\textsuperscript{2,*}
}
\IEEEauthorblockA{
\textsuperscript{1}University of New South Wales, 
\textsuperscript{2}University of Technology Sydney, 
\textsuperscript{3}Hong Kong Polytechnic University\\
\{yifan.zhu4, wenjie.zhang\}@unsw.edu.au, \{hanchen.wang, ying.zhang\}@uts.edu.au, alexander.zhou@polyu.edu.hk
}
}

% \author{\IEEEauthorblockN{1\textsuperscript{st} Given Name Surname}
% \IEEEauthorblockA{\textit{dept. name of organization (of Aff.)} \\
% \textit{name of organization (of Aff.)}\\
% City, Country \\
% email address or ORCID}
% \and
% \IEEEauthorblockN{2\textsuperscript{nd} Given Name Surname}
% \IEEEauthorblockA{\textit{dept. name of organization (of Aff.)} \\
% \textit{name of organization (of Aff.)}\\
% City, Country \\
% email address or ORCID}
% \and
% \IEEEauthorblockN{3\textsuperscript{rd} Given Name Surname}
% \IEEEauthorblockA{\textit{dept. name of organization (of Aff.)} \\
% \textit{name of organization (of Aff.)}\\
% City, Country \\
% email address or ORCID}
% \and
% \IEEEauthorblockN{4\textsuperscript{th} Given Name Surname}
% \IEEEauthorblockA{\textit{dept. name of organization (of Aff.)} \\
% \textit{name of organization (of Aff.)}\\
% City, Country \\
% email address or ORCID}
% \and
% \IEEEauthorblockN{5\textsuperscript{th} Given Name Surname}
% \IEEEauthorblockA{\textit{dept. name of organization (of Aff.)} \\
% \textit{name of organization (of Aff.)}\\
% City, Country \\
% email address or ORCID}
% \and
% \IEEEauthorblockN{6\textsuperscript{th} Given Name Surname}
% \IEEEauthorblockA{\textit{dept. name of organization (of Aff.)} \\
% \textit{name of organization (of Aff.)}\\
% City, Country \\
% email address or ORCID}
% }

\maketitle

\begingroup
\renewcommand{\thefootnote}{}
\footnotetext{
\textsuperscript{*}Corresponding Author.
}
\endgroup

\begin{abstract}
% *CRITICAL: Do Not Use Symbols, Special Characters, Footnotes, 
% or Math in Paper Title or Abstract.
    Searching and detecting communities in real-world graphs underpins a wide range of applications.
    Specifically, community search identifies a community for a given query, whereas community detection partitions the entire graph into multiple communities.
    Although they target different community analysis objectives, both tasks aim to uncover cohesive community structures and are complementary in real-world applications.
    Despite this connection, current learning-based solutions generally regard them as separate problems, necessitating task- and dataset-specific retraining.
    This separated learning paradigm limits the applicability and generalization ability of the existing models.
    To mitigate this limitation, we propose \model, a unified framework to solve both community search and detection tasks.
    \model centers on a Universal Graph Learning (UGL) stage that distills local and global community-aware knowledge from multiple source domains.
    Following UGL, the Domain-aware Specialization (DAS) stage supports on-the-fly adaptation to unseen graphs and tasks through transfer learning, avoiding costly retraining while keeping the model compact.
    Extensive experiments on both tasks across $16$ datasets and $22$ baselines have been conducted to ensure a comprehensive, rigorous, and fair evaluation.
    \model consistently outperforms all state-of-the-art baselines across all tasks while maintaining runtime efficiency.
\end{abstract}


\section{Introduction}

\begin{figure}[t]
  \centering
  \includegraphics[width=\linewidth]{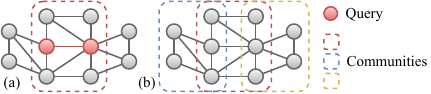}
  \vspace{-3mm}
  \caption{Overview of core tasks handled by \model. (a) Community Search (CS), (b) Community Detection (CD).}
  \vspace{-5mm}
  \label{fig:tasks}
\end{figure}

Graphs have demonstrated remarkable capability in modeling relational data across diverse domains, such as social networks~\cite{social}, e-commerce networks~\cite{amazon_coauthor_source}, and citation networks~\cite{planetoid_source}.
Communities~\cite{ProCom} are regarded as subgraphs whose nodes exhibit high structural and feature-wise cohesiveness, constituting a fundamental graph pattern for applications such as fraud detection~\cite{yu2024temporal} and recommendation systems~\cite{recommendation}.
As illustrated in \autoref{fig:tasks}, we focus on two major tasks: \textit{Community Search (CS)}~\cite{cs_definition}, which identifies a community based on a query, and \textit{Community Detection (CD)}~\cite{cd_definition}, which partitions the entire graph into multiple communities.
% ~\cite{cd_game}
Depending on whether nodes can belong to multiple communities, both tasks may involve either disjoint or overlapping communities.

\noindent \textbf{Existing Solutions.}
% \cite{TAS-Com}
Due to the importance of community-level tasks, numerous solutions~\cite{AdaptCS, CS_survey, CD_survey} have been proposed.
% \cite{InteractiveCS, s3gc, FoldCS}
% \cite{traditional1, traditional2, traditional3, traditional4}
Specifically, traditional algorithm-based methods~\cite{MkECS, W-CPM} primarily rely on rigid predefined constraints such as $k$-core~\cite{CST}, $k$-truss~\cite{CTC}, or $k$-clique~\cite{OCS}, which are rarely satisfied in real-world networks and render them overly \textit{sensitive to graph density}.
Moreover, they generally \textit{overlook the importance of node features}.
GNN-based methods~\cite{QD-GNN, SMN, CCGC, NOCD} mitigate the drawback by leveraging node features.
In particular, these methods commonly optimize a GNN model on a single task-specific dataset and then identify downstream communities by measuring representation similarities among nodes.
However, their dataset- and task-specific training strategy often results in \textit{slow convergence} and \textit{unstable training}.
% One potential solution is to involve a pre-training stage~\cite{GCL, SimGRACE} before task-specific tuning, which assists the model in initializing with generalizable representations.
% Nevertheless, on small graphs (e.g., Facebook~\cite{facebook_source}), these methods often \textit{fail to capture sufficient information} for effective model initialization, and on large graphs (e.g., Reddit~\cite{reddit_source}), it becomes challenging to extract comprehensive information from the entire graph due to \textit{computational and memory constraints}.
% Recently, graph foundation models (GFMs)~\cite{MDGFM, SAMGPT, RiemannGFM} have been proposed as a general solution to graph tasks with effective and robust representation learning.
% Nonetheless, these models are specifically designed for node- and graph-level classification tasks, overlooking the importance of \textit{cohesive information} and failing to adapt effectively to \textit{community-level tasks}.
% In addition, graphs often differ in both structure and feature distribution, while existing GFMs lack effective alignment and fusion mechanisms for transfer~\cite{GCOPE, MDGFM}, which could lead to \textit{information loss} or \textit{conflicts} during adaptation.
One potential solution is to introduce a graph pre-training~\cite{GCL, SimGRACE} stage prior to the task-specific learning, thereby providing generalizable representations for downstream datasets and tasks.
Existing graph pre-training methods commonly learn transferable representations through self-supervised objectives, such as contrasting augmented graph views or reconstructing observed structural relations.
Nevertheless, they typically lack explicit mechanisms to preserve community-aware information, resulting in \textit{limited support for community-level tasks}.
% Additionally, on small graphs (e.g., Facebook~\cite{facebook_source}), these pre-training methods may \textit{fail to capture sufficient community signals}, whereas on large graphs (e.g., Reddit~\cite{reddit_source}), extracting comprehensive graph information becomes \textit{computationally and memory-intensive}.
% Recently, graph foundation models (GFMs)~\cite{MDGFM, SAMGPT, RiemannGFM} have been proposed as more general solutions to graph learning.
% Nonetheless, most GFMs are still designed for node- and graph-level classification tasks, overlooking \textit{community-level cohesiveness} and lacking effective mechanisms to align and fuse \textit{transferable community-aware knowledge} across different graphs~\cite{GCOPE, MDGFM}, which may cause \textit{information loss} or \textit{conflicts} during the adaptation process.

\noindent \textbf{Motivation.}
Although CS and CD tasks differ in their output forms, they are intrinsically related at the community level. 
In particular, both tasks aim to identify cohesive and well-separated groups of nodes, and are governed by similar principles, including local compactness, boundary separability, and membership consistency.
Moreover, in practical graph analysis, CS and CD are often not isolated procedures but closely connected stages of community exploration.
CD provides a global view of community distribution by identifying overall community structures in the graph, while CS performs more fine-grained and query-specific community retrieval according to user requirements.
This sequential and hierarchical relationship indicates that CS and CD can be viewed as distinct downstream realizations of the same community-level knowledge, rather than as two independent problems.

Such community-level knowledge is not tied to a single graph or domain but repeatedly emerges in real-world networks, where nodes form cohesive groups based on structural, attribute, or semantic relations.
Graphs from different domains instantiate these shared principles in diverse ways, providing complementary information for learning robust community-aware representations.
This motivates multi-domain pre-training to learn reusable community-aware priors from diverse perspectives.
Following that, transfer learning can adapt these priors to different graphs, thereby supporting both CS and CD through lightweight, task-specific tuning.

As a consequence, a unified framework that leverages multi-domain knowledge for community-level tasks remains largely underexplored and faces two major challenges:

\textit{(1) Extracting Knowledge from Multiple Domains.}
Different source domains contain rich community-level priors, but these priors are often shaped by domain-specific characteristics.
Directly combining them into a single model may introduce interference, making it difficult to extract transferable knowledge while preserving community-aware information.

\textit{(2) Transferring Knowledge to Target Graphs.}
Although source domains provide informative priors, transferring such knowledge to a target graph remains challenging due to differences in graph distributions across source and target domains.
Such differences primarily arise from shifts in node feature spaces, in which feature semantics and dimensionality may vary substantially across domains.

\noindent \textbf{Our Solutions.}
To solve the aforementioned challenges, we introduce \model, a unified framework that jointly addresses CS and CD through cross-domain knowledge transfer.
Specifically, \model has a Universal Graph Learning (UGL) stage and a Domain-aware Specialization (DAS) stage.
To address Challenge (1), the UGL stage applies data augmentation both locally and globally to each source dataset and pre-trains dataset-specific models with community-aware objectives, enabling effective knowledge extraction while reducing cross-domain interference.
To address Challenge (2), the DAS stage introduces adaptation prompts and a feature projector for source--target alignment, and further uses anchor-based selection to identify representative nodes from both domains for Maximum Mean Discrepancy (MMD)-based optimization~\cite{mmd}.
Our major contributions can be summarized as:
\begin{itemize}[leftmargin=6mm]
    \item To the best of our knowledge, the proposed \model is the first unified framework targeting diverse community-level tasks, which jointly addresses CS and CD tasks for both disjoint and overlapping communities.
    \item We implement a task- and dataset-aware cohesive subgraph prompt and design a strategy to automatically generate multi-hop subgraphs, collectively ensuring local and global cohesiveness to support community-level tasks.
    \item We present a prompt-tuning framework with a task-aware fusion module for multi-domain knowledge integration.
    This framework enables knowledge integration by mitigating interference across domains and tasks.
    \item We conduct extensive experiments on $22$ baselines and $16$ real-world datasets, covering both disjoint and overlapping types, to comprehensively demonstrate the superior effectiveness of UniCom for both CS and CD tasks.
\end{itemize}

\section{Related Works}

\begin{figure*}[t]
  \centering
  \includegraphics[width=0.90\linewidth]{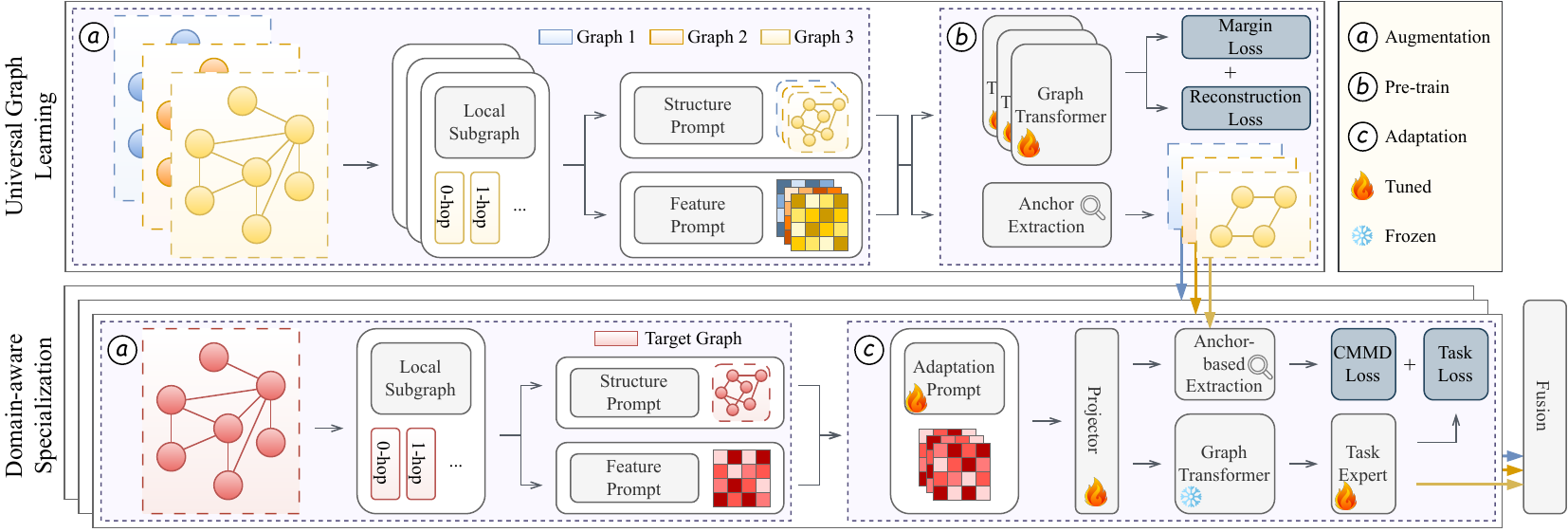}
  \vspace{0mm}
  \caption{Overview of \model. 1. In Universal Graph Learning (UGL), we pre-train multiple models in parallel using datasets from different domains. 2. In Domain-aware Specialization (DAS), the pre-trained models are frozen, while anchor nodes selected from pre-training datasets are transferred. 3. Finally, outputs from multiple models are fused for the downstream task.}
  \vspace{-3mm}
  \label{fig:overview}
\end{figure*}

\subsection{Community Search and Detection}

Community search is a fundamental task that discovers query-based subgraphs with high cohesiveness. Traditional methods~\cite{CST, CTC,  OCS} rely on predefined constraints but struggle to integrate multi-dimensional node attributes.
% \cite{traditional1, raditional2, raditional3, traditional4}
Inspired by recent advances in deep learning~\cite{pan2025hitec, pan2026universal, li2026fairness, xu2026c, tan2026skillzip, su2024bridging}, learning-based approaches have also been explored for community search~\cite{25sigmodCS, CommunityAF, CSFormer, ALICE}.
ICS-GNN~\cite{ICS-GNN} and SMN~\cite{SMN} search for query-dependent communities with user-defined size, whereas QD-GNN~\cite{QD-GNN} and COCLEP~\cite{COCLEP} predict community scores via GNNs followed by thresholding.
Additionally, CGNP~\cite{CGNP} and IACS~\cite{IACS} adopt meta-learning, and CommunityDF~\cite{CommunityDF} employs diffusion-based modeling.
Disjoint community detection assigns nodes to separate clusters.
% \cite{UniCD, DAG, spectral, CONVERT, DGCN, GUCD}
Specifically, CCGC~\cite{CCGC} adopts a two-stage framework with contrastive learning and pseudo-labels, DyFSS~\cite{DyFSS} combines multiple pre-training schemes, while FPGC~\cite{FPGC} leverages cluster-specific features via squeeze-and-excitation blocks.
Overlapping community detection allows nodes to exist in multiple communities.
NOCD~\cite{NOCD} employs a Bernoulli-Poisson decoder with GNNs, extended by DynaResGCN~\cite{DynaResGCN} with residual connections. UCoDe~\cite{UCoDe} designs a modularity-based contrastive learning, and SSGCAE~\cite{SSGCAE} further incorporates semi-supervised learning.
However, existing solutions only focus on the task-specific graph, overlooking knowledge and information from diverse domains, thus leading to suboptimal performance.

\subsection{Graph Transfer Learning}

Graph pre-training aims to equip GNNs with transferable structural and semantic knowledge via large-scale self-supervised learning, including generative~\cite{MGM, GPT-GNN} and contrastive methods~\cite{GCL, SimGRACE}. To bridge the gap between pre-training and downstream tasks, recent studies explore two adaptation paradigms: fine-tuning~\cite{GraphLoRA, GraphControl} and prompt learning~\cite{GPPT, GPF, GraphPrompt, All-in-one}, both of which freeze the pre-trained model and introduce additional learnable parameters. However, these solutions mainly focus on single-domain pre-training, lacking knowledge from multiple domains.
To address out-of-distribution generalization in cross-domain graph learning, researchers have built novel foundation models for textual graphs~\cite{OFA, GraphCLIP}.
% \cite{UniGraph}
Moreover, recent text-free graph foundation models~\cite{GCOPE, SAMGPT, MDGFM} consider multiple graphs from various domains and then generalize to downstream tasks.
% \cite{TrimannGFM}
Nevertheless, these models are designed for node- or graph-level classification tasks and neglect community-level cohesiveness, thus lacking adaptability to community-level tasks.

\section{Preliminaries}

Given an attributed graph \( G = (\mathcal{V}, \mathcal{E}) \), its adjacency matrix \( \bm{A} \in \{0,1\}^{|\mathcal{V}| \times |\mathcal{V}|} \) is defined such that \( \bm{A}_{u,v} = 1 \) if there is an edge between nodes \( u \) and \( v \), and \( \bm{A}_{u,v} = 0 \) otherwise. The feature matrix is $\bm{X} \in \mathbb{R}^{|\mathcal{V}| \times d}$, where each row $\bm{X}_v$ corresponds to the $d$-dimensional feature vector for node $v \in \mathcal{V}$.
% For clarity, we summarize the main notations used in this paper in \autoref{tab:notations}, including graph-related symbols, community and model variables, and internal mechanisms of the framework.
Following prior works~\cite{CS_survey2, CD_survey}, we consider homogeneous graphs as a well-established setting in this paper.

\subsection{Problem Statement}

In this work, we investigate cross-domain graph transfer learning for community-level tasks, including community search and community detection.

\begin{definition}[Community Search~\cite{cs_definition}]
Given a set of queries $\mathcal{Q}$, where each query contains one or more nodes, a graph $G = (\mathcal{V}, \mathcal{E})$, and a target size $r$, community search (CS) aims to identify, for each query, a query-specific cohesive subgraph of size $r$, as shown in \autoref{fig:tasks}a.
\end{definition}

\begin{definition}[Community Detection~\cite{cd_definition}]
Given a graph $G = (\mathcal{V}, \mathcal{E})$ and a predefined number of communities $K$, community detection (CD) aims to partition the node set into $K$ subsets $\{C_1, \dots, C_K\}$, where each $C_i \subseteq \mathcal{V}$ induces a cohesive subgraph, as illustrated in \autoref{fig:tasks}b.
\end{definition}

\begin{definition}[Disjoint and Overlapping Communities~\cite{SMN, NOCD}]
In disjoint communities, each node is assigned to one community, i.e., $C_i \cap C_j = \emptyset$ for $i \neq j$.
In contrast, overlapping communities allow a node to participate in multiple communities, i.e., $|\{i : v \in C_i\}| \geq 1$ for a node $v \in \mathcal{V}$.
\end{definition}

% Based on the above definitions, the CS task aims to identify the top-$r$ most relevant nodes for each query, where $r$ denotes the user-specified community size. Moreover, in the CD task, a node belongs to exactly one community in disjoint datasets and may belong to multiple communities in overlapping datasets.

In this paper, both CS and CD tasks are studied under disjoint and overlapping community settings.

\section{Methodology}

We propose \model, a unified framework for both community search (CS) and community detection (CD).
As illustrated in \autoref{fig:overview}, \model consists of two key components: Universal Graph Learning (UGL) and Domain-aware Specialization (DAS), both built upon the cohesiveness-aware augmentation that captures local and global information to learn community-aware patterns.
Finally, a fusion module aggregates multiple predictions for downstream tasks.

\subsection{Universal Graph Learning}

\noindent \textbf{Motivation.}
% Although community-level tasks exhibit diverse output forms, they are all rooted in patterns of community cohesiveness.
% Pre-training can provide useful community priors for downstream tasks by capturing such cohesiveness patterns.
% However, task-specific pre-training tends to capture limited dataset-specific patterns and requires re-training whenever a new task dataset is introduced.
% To learn more generalizable community priors, we propose UGL, a cross-domain, multi-dataset pre-training framework that learns from multiple source graph domains and transfers these priors to unseen task domains.
% To reduce interference among heterogeneous domains, UGL further maintains a separate domain-specific model for each pre-training dataset.
Although CS and CD exhibit different output formats, both tasks are fundamentally concerned with identifying groups of nodes that should be recognized as the same community.
Pre-training can provide community-aware priors for downstream tasks by capturing community-level information in each graph.
However, single-dataset pre-training often restricts such priors to a specific graph distribution, requiring separate pre-training for each new task dataset.
Therefore, UGL performs cross-domain and multi-dataset pre-training to learn community-aware priors from different perspectives.

\noindent \textbf{Local Augmentation.}
We first augment each input graph to enrich community-related information.
Neighborhoods provide locally cohesive information around a node, but the importance of different hops can vary substantially from node to node.
To better capture local information and adaptively determine the appropriate hop range, we introduce a conductance-based selection method that promotes both internal cohesion and external separability to support the augmentation locally.

\begin{definition}[Conductance~\cite{conductance1, conductance2}]
Let $G = (\mathcal{V}, \mathcal{E})$ be a graph and $C \subseteq \mathcal{V}$ be a community, the conductance of $C$ is:

{
\vspace{-0.5mm}
\small
\begin{equation}
    \operatorname{Cond}\left(G, C\right) = \frac{\operatorname{cut}(C, \overline{C})}{\min \left( \operatorname{deg}(C), \operatorname{deg}(\overline{C}) \right)},
\end{equation}
}

\noindent where $\overline{C} = \mathcal{V} \setminus C$ is the complement of $C$, $\operatorname{cut}\left( \cdot, \cdot \right)$ is the number of edges between $C$ and $\overline{C}$, and $\operatorname{deg}(\cdot)$ denotes the sum of degrees of all nodes within a given community.
\end{definition} 

We select the subgraph induced by multi-hop neighbors with the lowest conductance as the augmented local subgraph.
This strategy balances internal cohesion and external separability, enabling meaningful local information for each node.
We then perform message propagation over the selected subgraph to obtain the augmented local representation $\bm{X}^{\text{local}}$.

\noindent \textbf{Global Augmentation.}
Existing works~\cite{COCLEP} primarily rely on local neighborhood information, which captures short-range proximity but is insufficient to fully characterize community-level cohesiveness. In real-world cases, nodes within the same community may extend beyond multi-hop neighborhoods, where distant nodes with similar features or structural patterns can also be relevant. Therefore, we build cohesive subgraph prompts from both feature and structure perspectives.

Given a node $v$ in graph $G=(\mathcal{V},\mathcal{E})$ with node features $\bm{X}\in\mathbb{R}^{|\mathcal{V}|\times d}$,
applying feature-level partition (e.g., K-means~\cite{k-means}) to the node features yields
$\mathcal{C}^{\text{feat}}=\{C_1^{\text{feat}},\dots,C_{K_{\text{feat}}}^{\text{feat}}\}$, 
where node $v$ belongs to $C_n^{\text{feat}}$. 
Similarly, structure-level partition (e.g., Louvain) is applied to obtain 
$\mathcal{C}^{\text{strc}}=\{C_1^{\text{strc}},\dots,C_{K_{\text{strc}}}^{\text{strc}}\}$, 
where node $v$ is assigned to $C_n^{\text{strc}}$. 
Then the feature and structure prompts 
$\bm{P}_v^{\text{feat}}, \bm{P}_v^{\text{strc}}\in\mathbb{R}^d$ are constructed as:

{
\vspace{-1.5mm}
\small
\begin{equation}
\bm{P}_v^{\text{feat}} = \frac{1}{|C_n^{\text{feat}}|} \sum_{u\in C_n^{\text{feat}}} \bm{X}_u,
\quad
\bm{P}_v^{\text{strc}} = \frac{1}{|C_n^{\text{strc}}|} \sum_{u\in C_n^{\text{strc}}} \bm{X}_u.
\label{equ:prompts}
\end{equation}
}

\noindent Subsequently, the representation after local augmentation is prompted with cohesive subgraph prompts:

{
\small
\begin{equation}
\bm{X}^{\text{full}} = \operatorname{Concat} \left( \bm{X}^{\text{local}}, \bm{P}^{\text{feat}}, \bm{P}^{\text{strc}} \right) \in \mathbb{R}^{|\mathcal{V}| \times m \times d}.
\end{equation}
}

\noindent The prompted feature matrix is denoted as $\bm{X}^{\text{full}}$, where each node is associated with $m$ token vectors, incorporating both local and global cohesive information, thus providing comprehensive inputs for downstream tasks.

\begin{figure}[t]
  \centering
  \includegraphics[width=\linewidth]{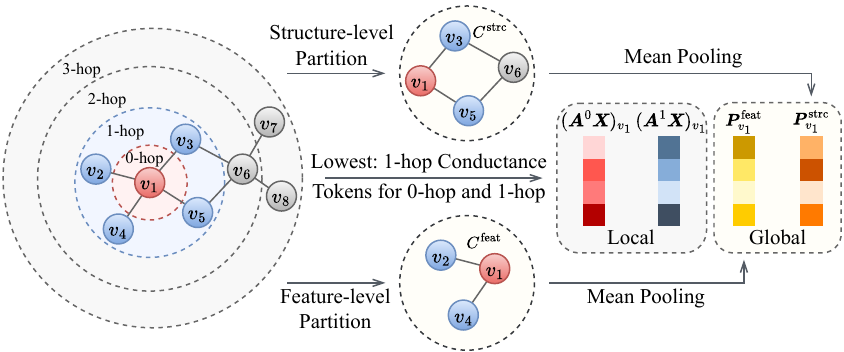}
  \vspace{-3mm}
  \caption{Local and Global Augmentation.}
  \vspace{-6mm}
  \label{fig:pre-process}
\end{figure}

\begin{example}
    Given the graph shown in \autoref{fig:pre-process}, for node $v_1$, its 1-hop neighborhood contains nodes $v_2$, $v_3$, $v_4$, and $v_5$. The conductance of the 1-hop local subgraph is thus $\frac{2}{4 + 1 + 1} = \frac{1}{3}$.
    Similarly, the conductance of the 2-hop local subgraph is $1$.
    The 1-hop neighborhood is selected since it has the lowest conductance.
    We then perform feature propagation to obtain the 0-hop and 1-hop feature tokens for node $v_1$, thus forming $\bm{X}_{v_1}^{\textnormal{local}}$.
    The feature- and structure-level partitions yield multiple communities, where $C^{\textnormal{feat}}$ and $C^{\textnormal{strc}}$ refer to the communities to which node $v_1$ belongs in each partition.
    Then $\bm{P}_{v_1}^{\textnormal{feat}}$ and $\bm{P}_{v_1}^{\textnormal{strc}}$ are constructed according to Equation~\ref{equ:prompts}.
\end{example}

\noindent \textbf{Community Encoder.}
To fully leverage the rich input with both local and global information, we employ a graph transformer~\cite{NAGphormer} to capture long-range dependencies and obtain expressive representations.
Specifically, given the augmented input $\bm{X}^\text{full}$, we then obtain the encoded node features as $\bm{H} = \operatorname{GT} \left( \bm{X}^\text{full}, \theta_g \right)$, where $\operatorname{GT}(\cdot, \theta_g)$ denotes the transformer encoder, and $\theta_g$ represents the learnable parameters tuned in the UGL stage while kept frozen in the DAS stage.
Each transformer block adopts a pre-norm architecture~\cite{LN} with residual connections, with the $l$-th block defined as:

{
    \small
    \begin{equation}
    \begin{aligned}
    \tilde{\bm{H}}^{(l)}_v
    &= \operatorname{MHA}\!\left(\operatorname{LN}\!\left(\bm{H}^{(l-1)}_v\right)\right)
     + \bm{H}^{(l-1)}_v, \\
    \bm{H}^{(l)}_v
    &= \operatorname{FFN}\!\left(\operatorname{LN}\!\left(\tilde{\bm{H}}^{(l)}_v\right)\right)
     + \tilde{\bm{H}}^{(l)}_v .
    \end{aligned}
    \end{equation}
}

\noindent Given the last-block output $\bm{H}^{(L)}$, we take the first token as the node-level representation $\bm{H}^{\text{node}}$ and mean-pool the remaining tokens to obtain the community-level representation $\bm{H}^{\text{com}}$.

\noindent \textbf{Optimization.}
For each source graph dataset, we first apply the cohesiveness-aware augmentation and then adopt the community encoder to generate representations.
To align node-level and corresponding community-level representations, we optimize the model with the following margin loss:

{
    \vspace{-1mm}
    \small
    \begin{equation}
        \mathcal{L}_{\text{mar}} \! = \! 
        \frac{1}{|\mathcal{V}|^2} \!\! \sum_{u,v \in \mathcal{V}}
        \!\!\! - \! \max\left( 
        \sigma \! \left( \bm{H}_u^{\text{node}} \bm{H}_u^{\text{com}} \right)
        \! - \! \sigma \! \left( \bm{H}_u^{\text{node}} \bm{H}_v^{\text{com}} \right) 
        \! + \! \epsilon, \, 0
        \right),
    \end{equation}
}

\noindent where $\sigma(\cdot)$ is the activation function and $\epsilon$ is the margin value.
In addition, based on the adjacency matrix $\bm{A}$, the reconstruction loss encourages similarity between connected nodes and discourages it between unconnected ones:

{
    \vspace{-1mm}
    \small
    \begin{equation}
        \mathcal{L}_{\text{rec}} \! = \!
        \frac{1}{|\mathcal{V}|^2} \!\! \sum_{u,v \in \mathcal{V}} \!
        (1 - \bm{A}_{u,v}) \left( \bm{H}_u^{\text{node}} \bm{H}_v^{\text{node}} \right)
        - \bm{A}_{u,v} \left( \bm{H}_u^{\text{node}} \bm{H}_v^{\text{node}} \right).
    \end{equation}
}

\noindent The overall pre-training objective comprises both the margin loss and the reconstruction loss: $\mathcal{L}_{\text{ugl}} = \mathcal{L}_{\text{mar}} + \alpha \mathcal{L}_{\text{rec}}$, where $\alpha > 0$ is the coefficient for balancing the two loss functions. Overall, the UGL stage equips domain-specific models with both local and global community information, providing a unified prior for downstream community-level tasks.

\noindent \textbf{Anchor Extraction.}
To support domain alignment in the DAS stage, we further derive a set of source domain anchors $\mathcal{B}^{\text{src}}$ for each pre-training graph. Specifically, the anchors are representative nodes from each source graph, identified using methods such as K-means or PageRank, thereby summarizing the key characteristics of their respective domains.

\subsection{Domain-aware Specialization}

\noindent \textbf{Motivation.}
The UGL stage provides community-aware priors from multiple source domains, while transferring such knowledge to unseen target graphs requires bridging the source--target domain gap.
Such a gap primarily arises from differences in structural patterns and feature semantics between the source and target graphs, thereby limiting the direct applicability of source-domain priors.
We therefore develop a transfer learning-based DAS stage that efficiently transfers multi-domain knowledge to target graphs for community-level tasks via adaptation prompts and a feature projector, both of which are optimized using community-aware objectives.

\noindent \textbf{Adaptation Prompt.}
To mitigate the distribution gap between source and target domains, we propose a lightweight graph prompt learning mechanism inspired by the ``pre-training and prompting'' paradigm, enabling cross-domain transfer with learnable prompts while keeping the pre-trained model frozen.
Before prompt construction, the target downstream dataset is processed with the same local and global augmentation strategies adopted in the UGL stage to generate $\bm{X}^\text{full}$, ensuring that both domains share a consistent augmented input format.
Then, a set of shared trainable basis prompt vectors $\{\bm{P}_i^{\text{basis}}\}_{i=1}^{N_p}$ is introduced, where $\bm{P}_i^{\text{basis}} \in \mathbb{R}^d$, and $N_p$ denotes the number of basis prompts~\cite{GPF}.
For each node $v$ and token position $t$, an adaptation prompt $\bm{P}_{v, t}^{\text{adapt}}$ is constructed as:

{
\vspace{-1mm}
\small
\begin{equation}
\bm{P}_{v, t}^{\text{adapt}} = \sum_{i=1}^{N_p} \omega_{v, t, i} \bm{P}_i^{\text{basis}},
\end{equation}
}

\noindent where $\omega_{v, t, i}$ is a learnable attention weight indicating the importance of the $i$-th prompt for node $v$ at token position $t$.
Then, all adaptation prompts are integrated with the augmented features $\bm{X}^\text{full}$ via element-wise addition to obtain the prompted $\bm{X}^{\text{adapt}}$.
Serving as a dual-channel bridge, the adaptation prompts align the feature distribution across domains.
% However, challenges arising from inconsistent feature dimensionality persist.

\noindent \textbf{Feature Projector.}
A feature projector is further introduced to align the inconsistent feature dimensions between source and target graphs in transfer learning, which is formulated as
$
\bm{Z} = \operatorname{Proj}\left(\bm{X}^{\text{adapt}}, \theta_p\right),
$
where $\operatorname{Proj}\left(\cdot, \theta_p\right)$ is a learnable function parameterized by $\theta_p$, for example, a multi-layer perceptron (MLP) in practice. Compared with traditional dimensionality reduction techniques such as Singular Value Decomposition (SVD), which overlook semantic consistency across dimensions, the proposed learnable projector preserves and aligns informative features during transfer.

\noindent \textbf{Task-specific Head.}
For both CS and CD tasks, after encoding the projected feature matrix $\bm{Z}$ via the community encoder, we concatenate $\bm{H}^{\text{node}}$ and $\bm{H}^{\text{com}}$ as the representations of all nodes, incorporating both node- and community-level information.
To identify the most relevant community of size $r$ for each query in the CS task, we introduce a cross-attention mechanism to capture interactions between the query and graph nodes. 
Specifically, attention scores are computed between representations of nodes in the query set and all potential nodes in the graph.
Then, top-$r$ nodes from the graph with the highest scores form the predicted community.
For the CD task, K-means clustering is employed as the task-specific head on the encoded representations to partition the graph into communities for a disjoint community dataset. In contrast, on overlapping datasets, we intend to produce a soft community affiliation matrix $\hat{\bm{Y}} \in \mathbb{R}^{|\mathcal{V}| \times K}$, where $\hat{\bm{Y}}_{v,i} \in [0, 1]$ denotes the probability that node $v$ belongs to community $C_i$. In practice, we employ a single MLP with an activation function (e.g., ReLU) as the decoder to project the embeddings to a target space matching the number of communities, thereby producing community assignment scores for each node.

\noindent \textbf{Anchor-based Extraction.}
Previous methods~\cite{MDGFM, GCOPE} either use a single task-driven objective for both domain adaptation and task learning, limiting alignment quality, or jointly bridge all source and target samples, resulting in high computational cost.
To efficiently optimize the adaptation prompt and projector, we adopt a hard-sampling strategy~\cite{Align-before-Fuse} on the target graph. Specifically, in each epoch, we select target nodes with the lowest cosine similarity between their projected features and the augmented features of source anchors $\mathcal{B}^{\text{src}}$ obtained in the UGL stage, and denote this set of target nodes as $\mathcal{B}^{\text{tar}}$.

\noindent \textbf{Alignment Optimization.}
Based on the selected nodes from both domains, we optimize adaptation prompts and the projector using Community Maximum Mean Discrepancy (CMMD), an MMD-inspired objective~\cite{mmd} that promotes domain alignment in transfer learning without labeled data:

{
    \vspace*{-2mm}
    \small
    \begin{align}
    \mathcal{L}_{\text{cmmd}} = \!\! \ 
    &\ \frac{1}{|\mathcal{B}^{\text{tar}}|^2} \!\! \sum_{u, u' \in \mathcal{B}^{\text{tar}}} \!\! \kappa(\bm{Z}^{\text{tar}}_u, \bm{Z}^{\text{tar}}_{u'}) 
    \! + \!  \frac{1}{|\mathcal{B}^{\text{src}}|^2} \!\! \sum_{v, v' \in \mathcal{B}^{\text{src}}} \!\! \kappa(\bm{X}^{\text{src}}_v, \bm{X}^{\text{src}}_{v'}) \notag \\
    &\ - \frac{2}{|\mathcal{B}^{\text{tar}}||\mathcal{B}^{\text{src}}|} \sum_{u \in \mathcal{B}^{\text{tar}}} \sum_{v \in \mathcal{B}^{\text{src}}} \kappa(\bm{Z}^{\text{tar}}_u, \bm{X}^{\text{src}}_v),
    \label{equ:loss_cmmd}
    \end{align}
}

\noindent where $\kappa(\cdot, \cdot)$ is the Gaussian kernel function.
The source representation $\bm{X}^{\text{src}}$ is obtained by mean pooling the cohesiveness-aware augmented node features in $\mathcal{B}^{\text{src}}$.
Similarly, the target representation $\bm{Z}^{\text{tar}}$ is obtained by mean pooling the projected node features in $\mathcal{B}^{\text{tar}}$.
This loss reduces the distribution discrepancy between the source and target domains.

\noindent \textbf{Community Search Optimization.} 
Beyond cross-domain alignment, we introduce task-specific training objectives to directly support each task.
For the CS task, given a training dataset with a query set $\mathcal{Q}$ and a label set $\mathcal{S}$ containing positive and negative nodes for each query, we optimize the trainable parameters using binary cross-entropy loss, guided by per-query supervision.
The loss function is defined as:

{
    \vspace{-2mm}
    \small
    \begin{equation}
    \mathcal{L}_{\text{cs}} = -\frac{1}{|\mathcal{Q}|} \sum_{q=1}^{|\mathcal{Q}|} 
    \frac{1}{|\mathcal{S}_q|} \sum_{v \in \mathcal{S}_q} 
    y_{q,v} \log\left(\pi_{q,v}\right)
    + \left(1 \! - \! y_{q,v}\right) \log\left(1 \! - \! \pi_{q,v}\right),
    \label{equ:loss_cs}
    \end{equation}
}

\noindent where $\pi_{q,v}$ is the prediction score for node $v$ in the $q$-th query, and $y_{q,v}$ is the corresponding ground-truth label from $\mathcal{S}$.

\noindent \textbf{Community Detection Optimization.} 
The CD task aims to assign community memberships to nodes in the graph. 
For disjoint datasets, we reuse the loss function from the pre-training phase for a few adaptation epochs.
Moreover, a pseudo-label-based refinement loss~\cite{CCGC} is employed, where high-quality labels are generated from nodes whose node- and community-level embeddings are most similar to each other.
Specifically, this refinement loss is formulated as follows:

{
    \vspace{-2mm}
    \small
    \begin{equation}
    \mathcal{L}_\text{dcd} \! = \!
    \frac{1}{K} \! \sum_{i=1}^{K} \! \sum_{v \in \mathcal{B}^{\text{pos}}_i} \! \left( 2 \! - \! 2 \cdot \text{sim} \left( \bm{H}_v^{\text{node}}, \bm{H}_v^{\text{com}} \right) \right)
    \! + \left\| \bm{S} \! - \! \operatorname{diag}(\bm{S}) \right\|_F^2,
    \label{equ:loss_dcd}
    \end{equation}
}

\noindent where $\mathcal{B}^{\text{pos}}_i$ denotes the subset of high-confidence nodes assigned to the cluster $i$, selected based on the embedding cosine similarity between representations $\bm{H}^{\text{node}}$ and $\bm{H}^{\text{com}}$.
$\bm{S} \in \mathbb{R}^{K \times K}$ is the inter-class similarity matrix computed between the mean embeddings (centers) of each cluster in the two views.
The term $\|\bm{S} - \operatorname{diag}(\bm{S})\|_F^2$ removes the diagonal elements and penalizes off-diagonal similarities, encouraging better separation between different communities. For overlapping datasets, we follow~\cite{NOCD} and minimize the Bernoulli–Poisson negative log-likelihood, aligning community memberships of connected nodes and separating unconnected pairs based on the soft assignment matrix $\hat{\bm{Y}}$. The loss is defined as:

{
    \vspace{-2mm}
    \small
    \begin{equation}
    \mathcal{L}_{\text{ocd}} \! = \!
    \frac{1}{|\mathcal{E}|} \!\! \sum_{(u,v)\in \mathcal{E}} 
    \!\!\! - \log\left(1 \! - \! \exp\left(-\epsilon \! - \! \hat{\bm{Y}}_u^\top \hat{\bm{Y}}_v\right)\right) 
    + \frac{1}{|\overline{\mathcal{E}}|} \sum_{(u,v)\in \overline{\mathcal{E}}} \!\! \hat{\bm{Y}}_u^\top \hat{\bm{Y}}_v,
    \label{equ:loss_ocd}
    \end{equation}
}

\noindent where $\epsilon$ is the small non-negative offset for likelihood stabilization.
Specifically, based on the assignment matrix $\hat{\bm{Y}}$, the first term increases the probability that connected node pairs belong to the same community, while the second decreases it for unconnected pairs, thereby ensuring structural consistency.

\noindent \textbf{Overall Objective Function.}
Building upon UGL, our DAS enables source-to-target transfer learning by jointly optimizing domain alignment and task-specific objectives.
The overall objective function is then defined as $\mathcal{L}_{\text{das}} = \mathcal{L}_{\text{task}} + \beta \mathcal{L}_{\text{cmmd}}$, where $\beta > 0$ is the trade-off parameter, and $\mathcal{L}_{\text{task}}$ denotes the corresponding task-specific objective, being $\mathcal{L}_{\text{cs}}$, $\mathcal{L}_{\text{dcd}}$, or $\mathcal{L}_{\text{ocd}}$.

\noindent \textbf{Multi-domain Knowledge Fusion.}
Finally, domain-specific structures and feature distributions in each source dataset encourage source-specific specialization in each model, causing predictions to focus on distinct perspectives.
As a consequence, relying on a single prediction reduces diversity and limits generalization to the target task.
To integrate these specializations and balance the complementary patterns captured by different pre-trained models, we apply a task-aware fusion approach within a unified framework that combines the strengths of different predictions.
For the CS task, we compute the average prediction score from all models for each node, and then select the top-$r$ nodes with the highest scores as the predicted community.
To leverage transferred knowledge for disjoint CD, we concatenate embeddings from all domains and apply the task-specific head for the final prediction. 
% For the overlapping CD task, we first align the multi-model predictions via consistent label mapping to avoid mismatch in community indices across different models,
For the overlapping CD task, we first align the multi-model predictions via consistent label mapping to avoid mismatched community indices across models, where the mapping is obtained by matching each model’s probability matrix to that of the first model using cosine similarity and the Hungarian algorithm.
Next, we average these prediction matrices and apply a thresholding operation to transform soft assignments into discrete community memberships.
% We further demonstrate the full DAS pipeline as presented in Algorithm~\ref{alg:das}.

\begin{lemma}[Fusion-Induced Risk Upper Bound]
For multi-domain knowledge fusion, the fused prediction for the CS task is no worse than the average individual expert with respect to the task objective.
The same upper-bound property also extends to the CD task under the corresponding objective.
\end{lemma}

\begin{proof}
In the CS task, DAS produces $N$ biased predictions $\{\pi_n\}_{n=1}^N$ and adopts a fusion mechanism
$\pi_{\text{out}}=\frac{1}{N}\sum_{n=1}^N \pi_n$.
Since the negative log-likelihood $\mathcal{L}_{\text{cs}}$ is strictly convex, Jensen's inequality gives
$\mathcal{R}(\pi_{\text{out}}) \le \frac{1}{N}\sum_{n=1}^N \mathcal{R}(\pi_n)$,
with strict inequality when predictions are not identical.
Thus, DAS fusion never increases the average risk and benefits from complementary domain-specific predictions.
The proof for disjoint CD follows from the additive squared Euclidean distance across views, while that for overlapping CD applies Jensen's inequality due to convexity in the similarity score.
\end{proof}

\subsection{Time Complexity Analysis}

\noindent \textbf{Graph Transformer.}
Each of the three projections in the graph transformer incurs a cost of $O(m d^2)$. 
Computing the query--key dot products and the attention-weighted values both require $O(m^2 d)$ time. 
Therefore, training a single model with $L$ layers for $t$ epochs over $|\mathcal{V}|$ nodes results in a total time complexity of
$O\left(t L |\mathcal{V}| (m d^2 + m^2 d)\right)$.

\noindent \textbf{Adaptation.}
% --------------- Adaptation Prompt and Feature Alignment ---------------
Computing adaptation prompts for $|\mathcal{V}|$ nodes costs $O(|\mathcal{V}| m N_p d)$, where $m$ is the number of tokens, $N_p$ is the number of basis prompts, and $d$ is the feature dimension.
Feature alignment with an $L$-layer MLP costs $O(L |\mathcal{V}| m d^2)$.
% ------------------------- Task Expert -------------------------
% ----- CS task -----

\noindent \textbf{Task-specific Head.}
The CS task employs a lightweight cross-attention mechanism over the query and all node embeddings, then the total search time is $O(|\mathcal{Q}|\,|\mathcal{V}|\,d)$ for $|\mathcal{Q}|$ queries.
% ----- DCD task -----
The disjoint CD employs K-means for graph partition with a time complexity of $O\left(|\mathcal{V}| |\mathcal{C}| d t\right)$ when iterated for $t$ steps.
% ----- OCD task -----
Overlapping CD employs a two-layer MLP to project node embeddings from hidden dimension $d$ to $|\mathcal{C}|$ communities, with a time complexity of $O\left(|\mathcal{V}| (d^2 + d|\mathcal{C}|)\right)$.

\section{Experiments}

\begin{table}[tb]
\centering
\caption{Statistics of Datasets.}
\vspace{0mm}
\renewcommand{\arraystretch}{1.00}
\setlength{\tabcolsep}{4pt}
\resizebox{0.95\linewidth}{!}{
\begin{tabular}{c|c|c|c|c|c|c|c}
\toprule
\rowcolor[gray]{0.9}
\textbf{Type} & \textbf{Dataset} & \textbf{\# N} & \textbf{\# E} & \textbf{\# C} & \textbf{\# F} & \textbf{OR} & \textbf{MLA} \\
\midrule
\multirow{5}{*}{\rotatebox{90}{\textbf{Pre-training}}}
& Instagram     & 11,339      & 155,349  & 2  & 500   & -         & -  \\
& WikiCS        & 11,701      & 431,726  & 10 & 300   & -         & -  \\
& Computers     & 13,752      & 491,722  & 10 & 767   & -         & -  \\
& DBLP          & 17,716      & 105,734  & 4  & 1,639 & -         & -  \\
& CoCS          & 18,333      & 163,788  & 15 & 6,805 & -         & -  \\
\midrule
\multirow{6}{*}{\rotatebox{90}{\textbf{Disjoint}}}
& Cora          & 2,708       & 10,556   & 7  & 1,433 & -         & -  \\
& CiteSeer      & 3,327       & 9,104    & 6  & 3,703 & -         & -  \\
& Photo         & 7,650       & 238,162  & 8  & 745   & -         & -  \\
& PubMed        & 19,717      & 88,648   & 3  & 500   & -         & -  \\
& Reddit        & 232,965     & 114M     & 41 & 602   & -         & -  \\
& Products      & 2,449,029   & 124M     & 47 & 100   & -         & -  \\
\midrule
\multirow{5}{*}{\rotatebox{90}{\textbf{Overlapping}}}
& FB1912        & 756         & 60,805   & 46 & 480   & 34.52     & 6  \\
& FB1684        & 793         & 28,840   & 17 & 319   & 0.76      & 3  \\
& FB107         & 1,046       & 54,543   & 9  & 576   & 1.82      & 3  \\
& ENG           & 14,927      & 98,610   & 16 & 4,839 & 27.20     & 12 \\
& CS            & 21,957      & 193,500  & 18 & 7,793 & 27.54     & 13 \\
\bottomrule
\end{tabular}
}
\vspace{-2mm}
\label{tab:datasets}
\end{table}

\begin{table*}[tb]
\setlength{\tabcolsep}{6.5pt}
\centering
\caption{Performance of Community Search.}
\vspace{-1mm}
\renewcommand{\arraystretch}{0.90}
\begin{tabular}{c|ccccccccccc|c}
\toprule
% \rowcolor[gray]{0.9}
\textbf{Methods} & \textbf{Cora} & \textbf{CiteSeer} & \textbf{PubMed} & \textbf{Photo} & \textbf{Reddit} & \textbf{Products}
& \textbf{FB107} & \textbf{FB1684} & \textbf{FB1912} & \textbf{CS} & \textbf{ENG}
& \textbf{Avg +/-} \\
\midrule
CST      & 43.39 & 38.61 & 69.46 & 70.94 & 44.33 & 67.39 & 54.01 & 73.15 & 55.31 & 38.90 & 33.62 & -31.34\% \\
CTC      & 33.62 & 32.22 & 56.42 & 63.60 & 15.52 & 61.36 & 36.66 & 57.93 & 50.75 & 32.81 & 29.94 & -42.10\% \\
OCS      & 46.92 & 35.33 & 56.55 & 23.09 & -     & -     & 30.35 & 40.71 & 49.49 & 26.99 & 50.60 & -45.28\% \\
MkECS    & 44.50 & 38.32 & 69.21 & 71.07 & 41.72 & 57.76 & 53.90 & 73.24 & 55.07 & 35.29 & 31.69 & -32.92\% \\
\midrule
ICS-GNN  & 60.33 & 52.88 & 89.61 & 93.52 & 73.87 & 57.74 & 70.69 & 90.03 & 78.65 & 40.95 & 42.77 & -16.62\% \\
QD-GNN   & 65.95 & 52.11 & 60.41 & 25.11 & -     & -     & \underline{77.44} & 81.05 & 80.16 & 52.54 & 61.73 & -23.45\% \\
COCLEP   & 80.31 & 44.98 & 90.23 & \textbf{94.88} & -     & -     & 71.30 & \underline{91.54} & \underline{81.42} & 66.24 & 69.88 & -8.53\% \\
SMN      & \underline{84.75} & \underline{77.27} & 87.02 & 91.33 & \underline{84.97} & \underline{71.42} & 70.07 & 83.87 & 76.98 & \underline{72.66} & \underline{76.91} & -5.15\% \\
\midrule
GCOPE    & 83.49 & 75.34 & 90.93 & 94.32 & -    & -    & 71.62 & 90.36 & 78.84 & 65.94 & 69.33 & -5.27\% \\
SAMGPT   & 80.32 & 77.06 & \underline{91.13} & 93.76 & -    & -    & 72.71 & 90.72 & 79.65 & 71.08 & 70.99 & -4.46\% \\
MDGFM    & 75.66 & 71.14 & 80.23 & 93.51 & -    & -    & 72.58 & 85.91 & 78.11 & 65.37 & 64.56 & -8.95\% \\
\midrule
\model   & \textbf{93.83} & \textbf{77.71} & \textbf{94.51} & \underline{94.45} & \textbf{89.72} & \textbf{76.57} 
         & \textbf{81.00} & \textbf{92.21} & \textbf{81.63} & \textbf{74.85} & \textbf{77.40} & - \\
\bottomrule
\end{tabular}
\vspace{0mm}
\label{tab:performance_cs}
\end{table*}

\begin{table*}[tb]
\centering
\caption{Performance of Community Detection.}
\vspace{-1mm}
\renewcommand{\arraystretch}{0.90}
\setlength{\tabcolsep}{7.5pt}
\begin{tabular}{c|ccc|c||c|ccccc|c}
\toprule
% \rowcolor[gray]{0.9}
\textbf{Methods} & \textbf{Cora} & \textbf{Photo} & \textbf{Reddit} & \textbf{Avg +/-}
& \textbf{Methods} & \textbf{FB107} & \textbf{FB1684} & \textbf{FB1912} & \textbf{CS} & \textbf{ENG} & \textbf{Avg +/-} \\
\midrule
AE         & 47.58 & 27.63 & 16.30 & -43.62\% 
           & W-CPM      & 8.43 & 22.22 & 12.37 & - & - & -19.21\% \\
DeepWalk   & 62.49 & 72.21 & \underline{73.13} & -4.85\% 
           & BigCLAM    & 14.23 & 33.99 & 33.71 & - & - & -6.24\% \\
\midrule
CCGC       & 58.56 & 65.68 & - & -11.78\% 
           & NOCD       & 11.85 & 33.09 & 39.32 & \underline{46.09} & \underline{39.75} & -3.08\% \\
DyFSS      & 67.34 & \underline{73.19} & - & -3.64\% 
           & DynaResGCN & \underline{15.69} & \underline{40.84} & \underline{40.60} & 41.60 & 37.51 & -1.85\% \\
FPGC       & \underline{73.10} & 69.45 & - & -2.63\% 
           & SSGCAE     & 15.07 & 37.05 & 25.56 & 31.23 & 30.43 & -9.23\% \\
UCoDe      & 62.39 & 47.72 & - & -18.85\% 
           & UCoDe      & 12.41 & 36.83 & 35.64 & \textbf{46.16} & 33.63 & -4.17\% \\
\midrule
\model     & \textbf{73.43} & \textbf{74.38} & \textbf{74.57} & - 
           & \model     & \textbf{17.34} & \textbf{42.21} & \textbf{41.11} & 44.26 & \textbf{40.59} & - \\
\bottomrule
\end{tabular}
\vspace{0mm}
\label{tab:performance_cd}
\end{table*}

In this section, we report the results of comprehensive experiments using $16$ datasets from diverse domains and $22$ competitive baselines, covering both CS and CD tasks, to demonstrate the effectiveness and efficiency of \model.

\subsection{Experimental Setup}

\noindent \textbf{Datasets.}
Evaluation of \model is conducted on $11$ real-world datasets spanning citation networks~\cite{planetoid_source, amazon_coauthor_source}, social networks~\cite{reddit_source, facebook_source}, e-commerce networks~\cite{amazon_coauthor_source, products_source}, and academic networks~\cite{NOCD}, including $6$ disjoint and $5$ overlapping datasets.
Additionally, $5$ datasets from diverse domains~\cite{dblp_source, wikics_source, amazon_coauthor_source, instagram_source} are adopted as pre-training sources. 
Detailed statistics, including the numbers of nodes, edges, and communities, along with the node feature dimensions, for all datasets are provided in \autoref{tab:datasets}.
Following SMN~\cite{SMN}, we report Overlap Rates (OR) and Max Label Affiliations (MLA) on overlapping datasets to quantify multi-community membership.

\noindent \textbf{Data Splitting.}
For the CS task, \model handles single- and multi-node queries ($1$–$3$ nodes). For the training set, we sample $20$ queries per community on disjoint datasets, $30$ random queries on Facebook datasets, and use $5\%$ of nodes as queries on MAG datasets.
For each training query, we use $3$ positive nodes from the ground-truth community and $3$ negative nodes outside it as weak supervision signals.
We set $100$ test queries per dataset.
For the CD task, following CCGC~\cite{CCGC} and NOCD~\cite{NOCD}, we adopt the unsupervised setting.

\noindent \textbf{Baselines.}
We select a wide range of baselines for each task to comprehensively evaluate the performance and efficiency of \model.
For the \textbf{CS} task, we compare \model against
$4$ \textbf{algorithm-based methods}: CST~\cite{CST}, CTC~\cite{CTC}, OCS~\cite{OCS}, and MkECS~\cite{MkECS}, and
$4$ state-of-the-art \textbf{GNN-based baselines}: ICS-GNN~\cite{ICS-GNN}, QD-GNN~\cite{QD-GNN}, COCLEP~\cite{COCLEP}, and SMN~\cite{SMN}.
By adapting the task output modules, we further include $3$ \textbf{graph foundation models (GFMs)}: GCOPE~\cite{GCOPE}, SAMGPT~\cite{SAMGPT}, and MDGFM~\cite{MDGFM}.
The evaluation of the \textbf{disjoint CD} task includes $2$ \textbf{traditional methods}: AutoEncoder (AE)~\cite{AE} and DeepWalk~\cite{DeepWalk}, as well as $4$ \textbf{GNN-based models}: CCGC~\cite{CCGC}, UCoDe~\cite{UCoDe}, DyFSS~\cite{DyFSS}, and FPGC~\cite{FPGC}.
In addition, for the \textbf{overlapping CD} task, we adopt $2$ \textbf{traditional baselines}: W-CPM~\cite{W-CPM} and BigCLAM~\cite{BigCLAM}, along with $4$ \textbf{GNN-based models}: NOCD~\cite{NOCD}, DynaResGCN~\cite{DynaResGCN}, UCoDe~\cite{UCoDe}, and SSGCAE~\cite{SSGCAE}.
Since the GFMs are tailored for supervised learning tasks, we exclude them from the evaluation of the unsupervised CD task.

\noindent \textbf{Evaluation Metrics.}
We evaluate CS and disjoint CD using F1-score (F1), following prior works~\cite{SMN, CCGC}.
For overlapping CD, we instead adopt Overlapping Normalized Mutual Information (ONMI)~\cite{NOCD}, since each node may belong to multiple communities, making F1 unsuitable.
To ensure fair and robust evaluation, we run each experiment $5$ times and report the average performance to reduce randomness.

\noindent \textbf{Implementation Details.}
For the UGL phase, we pre-train models with a learning rate of $0.001$ for $100$ epochs with early stopping, and set $\alpha$ to $0.1$.
We adopt K-means with $10$ clusters for anchor extraction.
For the DAS phase, the learning rate is in $\{0.0001, 0.001, 0.01\}$ and the coefficient $\beta$ is in $\{0.01, 0.1, 1\}$. 
The community encoder uses a hidden size of $512$, $8$ attention heads, and a dropout rate of $0.1$. 
We adopt a $3$-dimensional positional encoding for all datasets, except Reddit, Products, CS, and ENG, which use a $10$-dimensional positional encoding. 
We set the maximum conductance augmentation hop to $5$. 
The batch size equals the number of nodes and is capped at $4000$ when out-of-memory (OOM) occurs.
Hyperparameter settings for baselines follow the search spaces reported in their original papers.
All experiments are conducted on a server equipped with an Intel Xeon 6248R CPU (512 GB RAM) and NVIDIA RTX A5000 GPUs, providing a consistent computational setup across all methods.
The source code and data are available at: \url{https://github.com/Yifan-Andy/UniCom}.

\begin{figure*}[tb]
    \centering
    \includegraphics[width=0.936\linewidth]{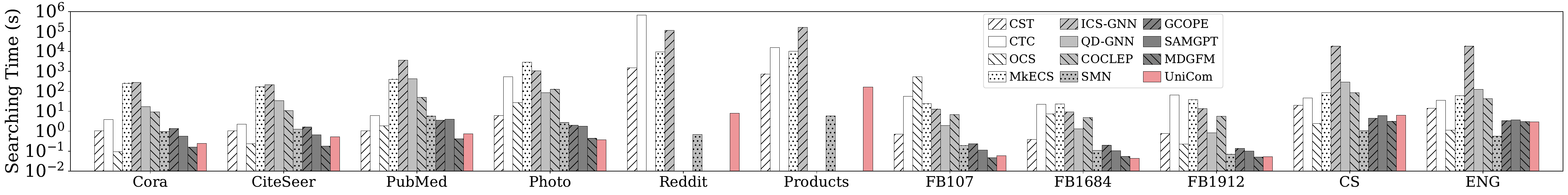}
    \vspace{-1mm}
    \caption{Total Online Searching Time of Community Search (100 Queries).}
    \vspace{-3mm}
    \label{fig:cs_time_online}
\end{figure*}

\begin{figure*}[t]
\centering
\begin{subfigure}[b]{0.473\linewidth}
  \centering
  \includegraphics[width=\linewidth]{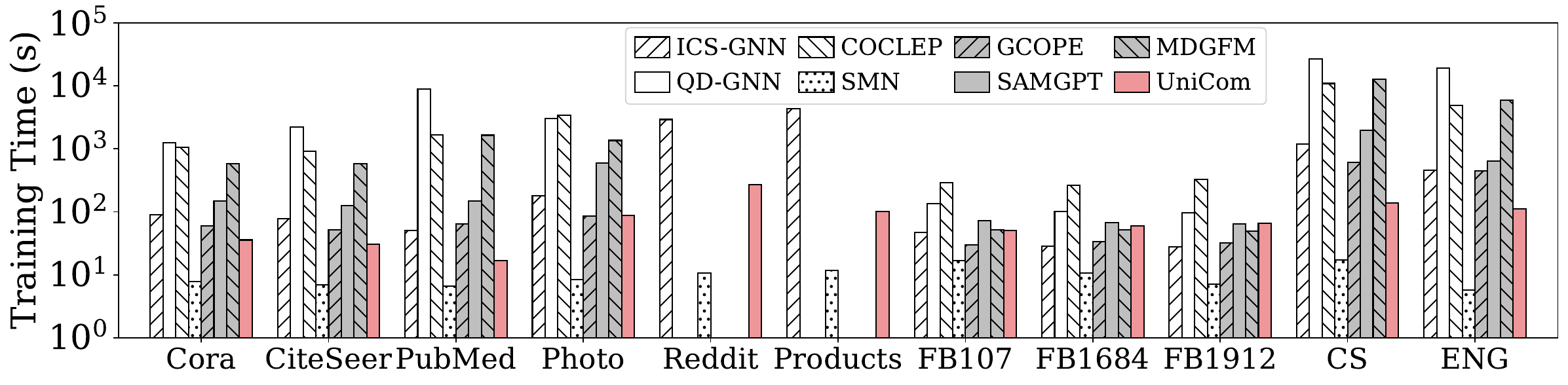}
  \vspace{-6mm}
  \caption{CS Training Time.}
  \label{fig:cs_time_offline}
\end{subfigure}\hfill
\begin{subfigure}[b]{0.2355\linewidth}
  \centering
  \includegraphics[width=\linewidth]{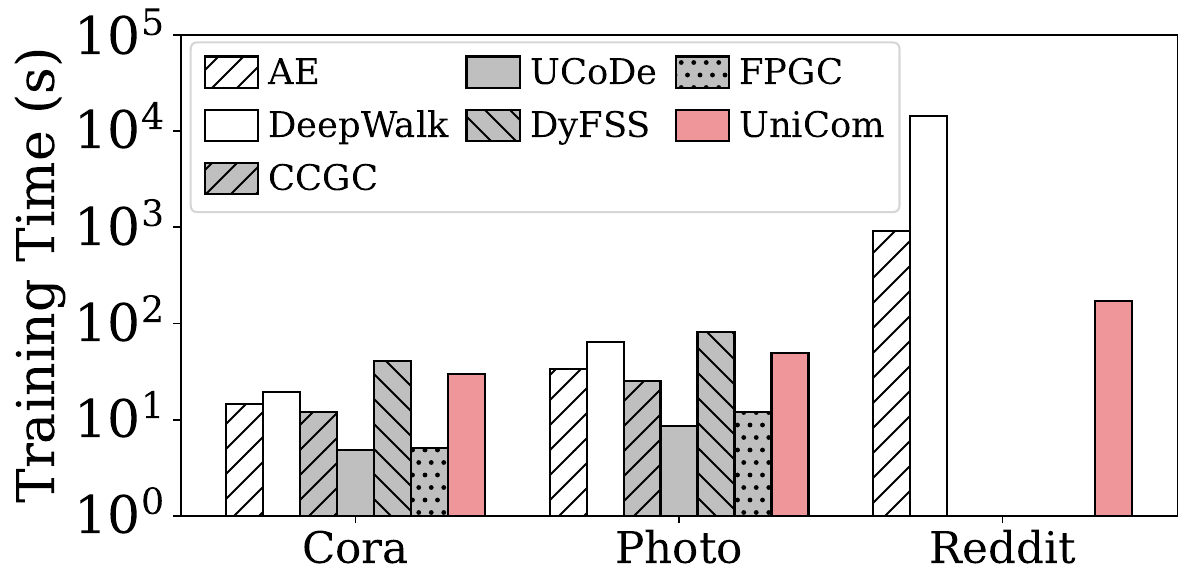}
  \vspace{-6mm}
  \caption{Disjoint CD Training Time.}
  \label{fig:cd_time_dcd}
\end{subfigure}\hfill
\begin{subfigure}[b]{0.280\linewidth}
  \centering
  \includegraphics[width=\linewidth]{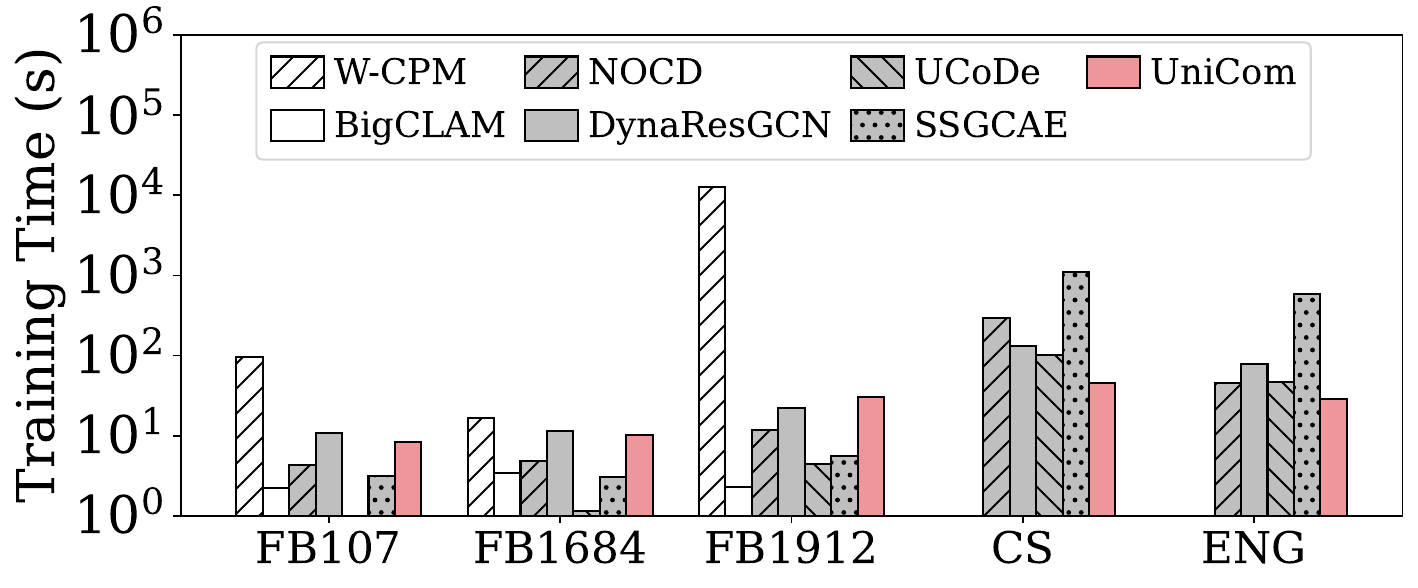}
  \vspace{-6mm}
  \caption{Overlapping CD Training Time.}
  \label{fig:cd_time_ocd}
\end{subfigure}

\vspace{0mm}
\caption{Training Time of Community Search and Detection.}
\vspace{-3mm}
\end{figure*}

\subsection{Effectiveness Evaluation}
We evaluate the effectiveness of \model on $11$ downstream datasets for CS and CD tasks and demonstrate the results in \autoref{tab:performance_cs} and \autoref{tab:performance_cd}.
The best results are highlighted in \textbf{bold}, while the second-best ones are \underline{underlined}.
Missing values denote OOM or runs that fail to complete within $7$ days.

\noindent \textit{Community Search.}
The performance comparison of \model against $11$ baselines is reported in \autoref{tab:performance_cs}.
The results show that \model outperforms all baselines on most datasets and ranks second on the rest. Moreover, the lightweight design enables \model to scale to large datasets without incurring OOM issues.
Compared with $4$ \textbf{algorithm-based methods} that overlook the pivotal role of node features, \model incorporates both structure- and feature-level information and outperforms CST (the best algorithm-based baseline) by $31.34\%$.
Compared with \textbf{GNN-based baselines}, \model achieves an average improvement of at least $5.15\%$, evidencing the importance of multi-domain knowledge for CS.
Additionally, \model outperforms all \textbf{GFMs} on the CS task across all datasets, with an average improvement of over $4.46\%$, highlighting the effectiveness of cohesive information.

\noindent \textit{Community Detection.}
We report the CD performance of \model in \autoref{tab:performance_cd}. Across $3$ benchmark datasets for disjoint CD, \model consistently achieves the highest F1-score, showing an average improvement of over $2.63\%$.
Notably, existing state-of-the-art \textbf{GNN-based models} often encounter OOM issues when applied to large-scale datasets such as Reddit.
In contrast, \model is explicitly designed for scalability, enabling adaptation on large graphs.
On the $5$ benchmark datasets with overlapping communities,
\model surpasses all baseline methods on $4$ out of $5$ datasets, with an average improvement ranging from $1.85\%$ to $19.21\%$.

Overall, \model consistently outperforms existing methods across all tasks, which often lack multi-dataset knowledge or cohesiveness modeling. 
This highlights the benefits of a unified framework that enables consistent community representation learning across diverse tasks and domains.

\subsection{Efficiency Evaluation}

\noindent \textbf{Searching Time.}
\autoref{fig:cs_time_online} shows the total online searching time of $100$ queries for all baselines and \model. Traditional methods generally exhibit lower search times on sparse graphs due to their non-learning nature. However, on graphs with high average degrees (e.g., Reddit), they suffer from a dramatic escalation in search time.
Compared with GNN-based methods and GFMs, \model generally has the top-$3$ fastest searching time on most datasets.
Notably, \model achieves significantly better efficiency than ICS-GNN.
Moreover, it is on average 86.55$\times$ and 30.24$\times$ faster than QD-GNN and COCLEP, respectively.
Compared with SMN and GFMs, \model also exhibits comparable or even higher efficiency.

% We further report the offline training time for the CS task and the model training time for the CD task in \autoref{fig:combined_time}

% % and provide time complexity analysis in \autoref{app:time_complexity}.

% Moreover, \model exhibits notable space efficiency, with significantly fewer tunable parameters compared to prior GNN-based approaches in both CS and CD tasks, which is additionally discussed in \autoref{app:additional_experiments}.

\noindent \textbf{Training Time.}
\autoref{fig:cs_time_offline} compares the CS training time of \model with baselines.
Since traditional CS methods require no offline training, we report training time only for GNN-based approaches and GFMs.
Although \model may require multiple offline training runs depending on the number of pre-training datasets used, these runs can be executed in parallel on multiple GPUs, resulting in high efficiency.
In general, \model achieves substantially shorter training time than most selected baselines, including ICS-GNN, QD-GNN, COCLEP, and all GFMs.
Based on the results in \autoref{fig:cd_time_dcd} and \autoref{fig:cd_time_ocd}, \model demonstrates highly efficient training in both disjoint and overlapping CD tasks.
Notably, on the Reddit dataset, \model is the only GNN-based method that completes CD without encountering OOM errors, all while keeping the training time within a practical range.
Total training time for overlapping CD shows that \model exhibits stable efficiency across $5$ overlapping datasets, with particularly minimal training time on large MAG academic datasets.

\subsection{Ablation Studies}

% In this section, we evaluate the contribution of each component in \model. 
% As illustrated in \autoref{tab:ablation_studies}, 
We conduct ablation studies by comparing the performance of \model and $4$ variants on both CS and CD tasks across disjoint and overlapping datasets to evaluate the contribution of each component.
% Specifically, ``w/o Pre-train'' refers to removing the multi-domain pre-training stage, where the model is directly fine-tuned for downstream tasks. 
Specifically, ``w/o Pre-train'' is a variant that replaces the \textit{multi-domain pre-training} with task-specific fine-tuning.
``w/o Fusion'' is a variant that removes the \textit{multi-domain knowledge fusion} module and is pre-trained solely on the DBLP dataset. 
Additionally, ``w/o Coh Prompt'' and ``w/o Adp Prompt'' indicate the removal of the \textit{cohesive subgraph prompt} and \textit{adaptation prompt}, respectively. 
% Notably, the variant without the adaptation prompt makes the pre-trained model fully tunable. 
The experiments are conducted on 6 datasets covering different tasks to assess the effectiveness of each component comprehensively.

The results are illustrated in \autoref{tab:ablation_studies}.
Overall, the full model consistently attains superior and robust performance across all tasks, with each component contributing complementarily. 
% The pre-training and adaptation prompt modules yield the most significant performance gain, highlighting the crucial role of knowledge transfer in subgraph-level tasks.
% The effect is particularly evident in CD tasks, where transferred knowledge compensates for the absence of supervision.
% These observations further prove the necessity of building a unified model for CS and CD.
The pre-training and adaptation prompt modules yield the largest performance gains, highlighting the importance of knowledge transfer in subgraph-level tasks.
The effectiveness is most evident in the CD task, where transferred knowledge compensates for the absence of ground-truth labels.
These observations further support the benefit of building a unified model for both CS and CD.
Moreover, the fusion module and cohesive subgraph prompt also offer stable benefits across diverse tasks, underscoring the value of cross-domain knowledge integration and the efficacy of transfer learning design for community-level tasks across various datasets.

\subsection{Hyper-parameter Analysis}

In this section, we systematically evaluate the performance of \model under various configurations, including the trade-off parameter $\beta$, the number of basis prompts for adaptation, and the number of pre-training graphs adopted.
% These experiments aim to provide a comprehensive analysis of how each factor influences the model’s effectiveness and to illuminate the robustness and applicability of the proposed approach.

\begin{figure}[t]
\vspace{1mm}
\centering

% -------- Row 1 --------
\begin{subfigure}[b]{0.325\linewidth}
  \centering
  \includegraphics[width=\linewidth]{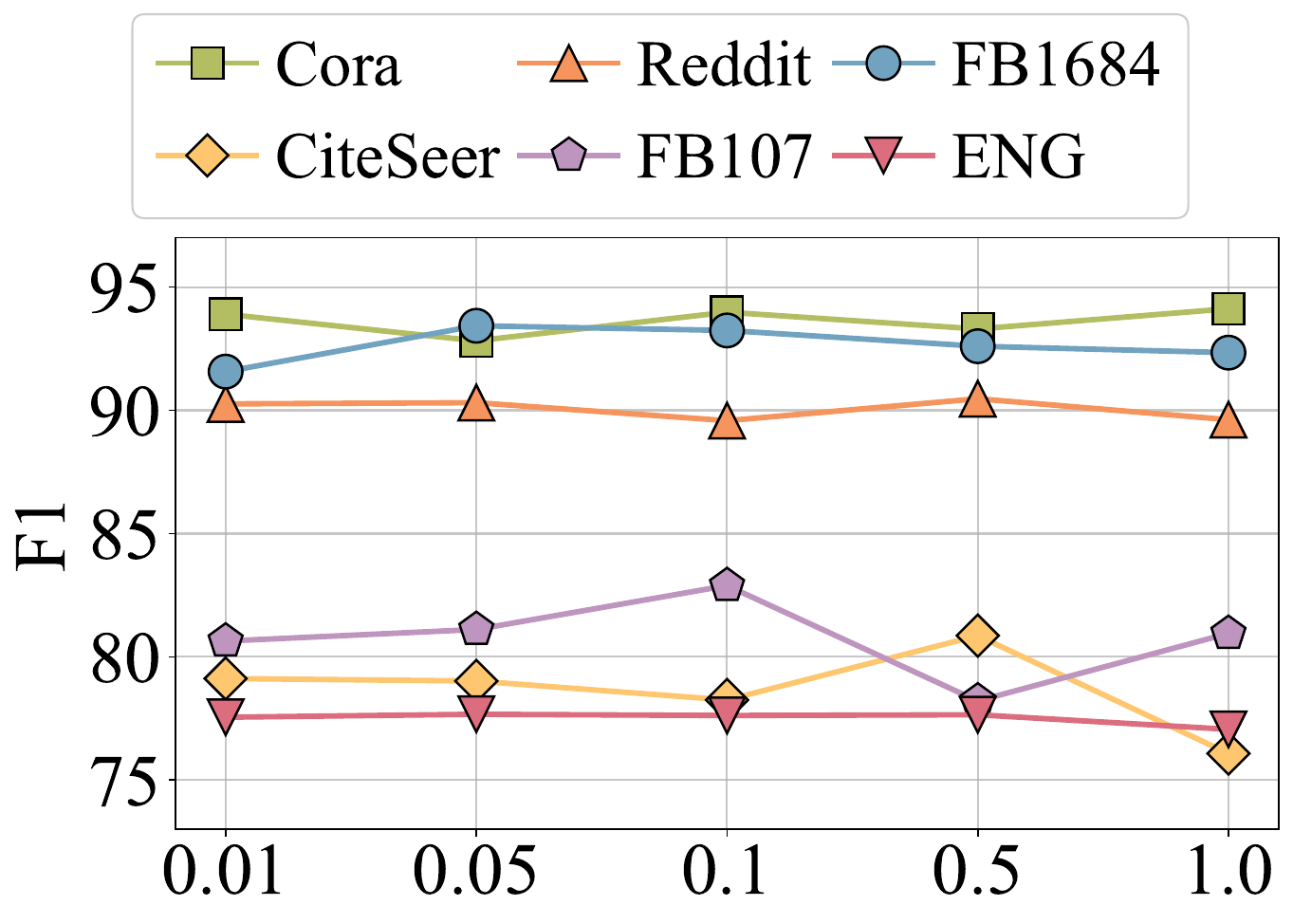}
  \vspace{-5.5mm}
  \caption{CS -- $\beta$.}
  \vspace{1mm}
  \label{fig:alpha_cs}
\end{subfigure}\hfill
\begin{subfigure}[b]{0.325\linewidth}
  \centering
  \includegraphics[width=\linewidth]{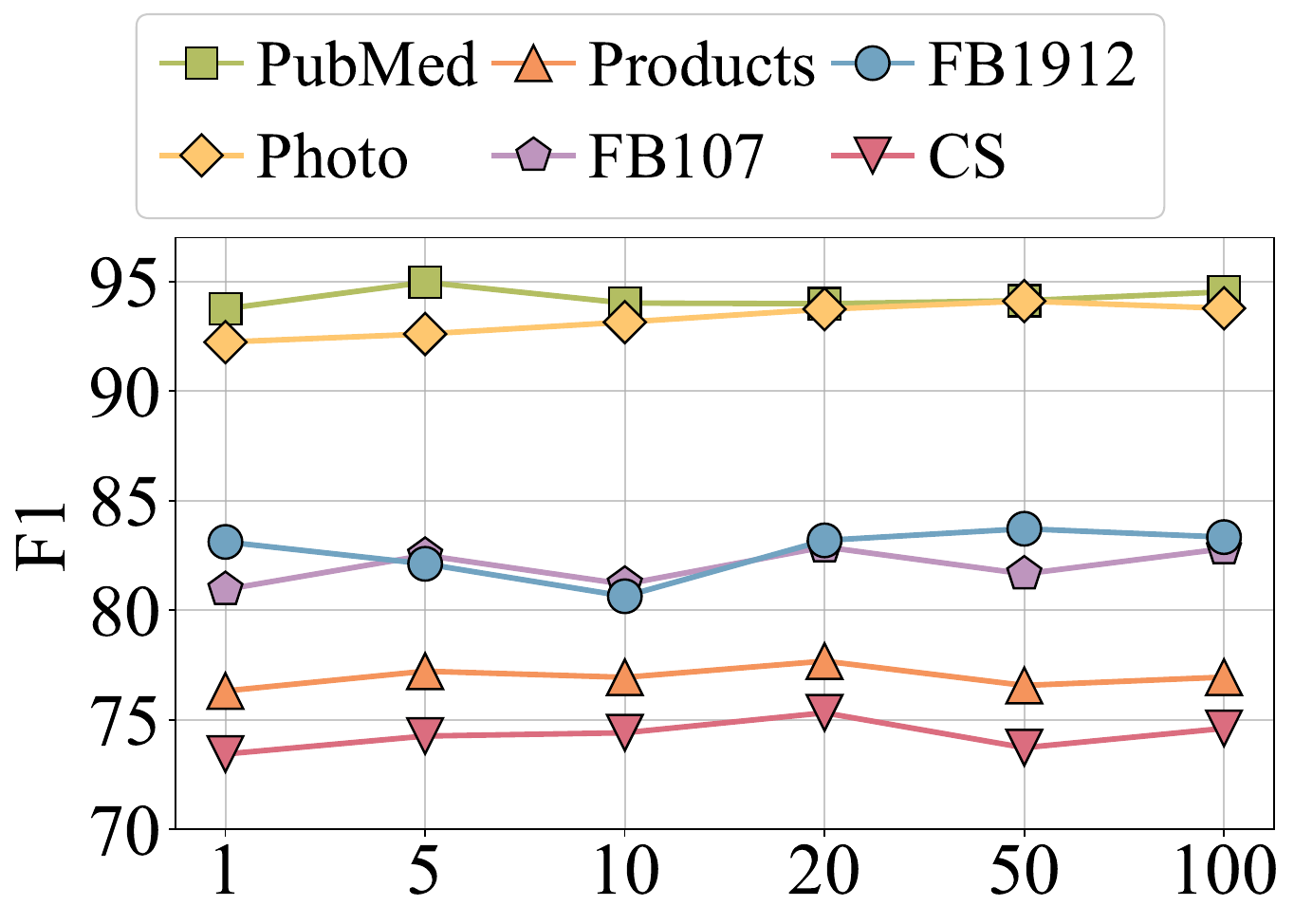}
  \vspace{-5.5mm}
  \caption{CS -- \# Prompts.}
  \vspace{1mm}
  \label{fig:n_prompt_cs}
\end{subfigure}\hfill
\begin{subfigure}[b]{0.325\linewidth}
  \centering
  \includegraphics[width=\linewidth]{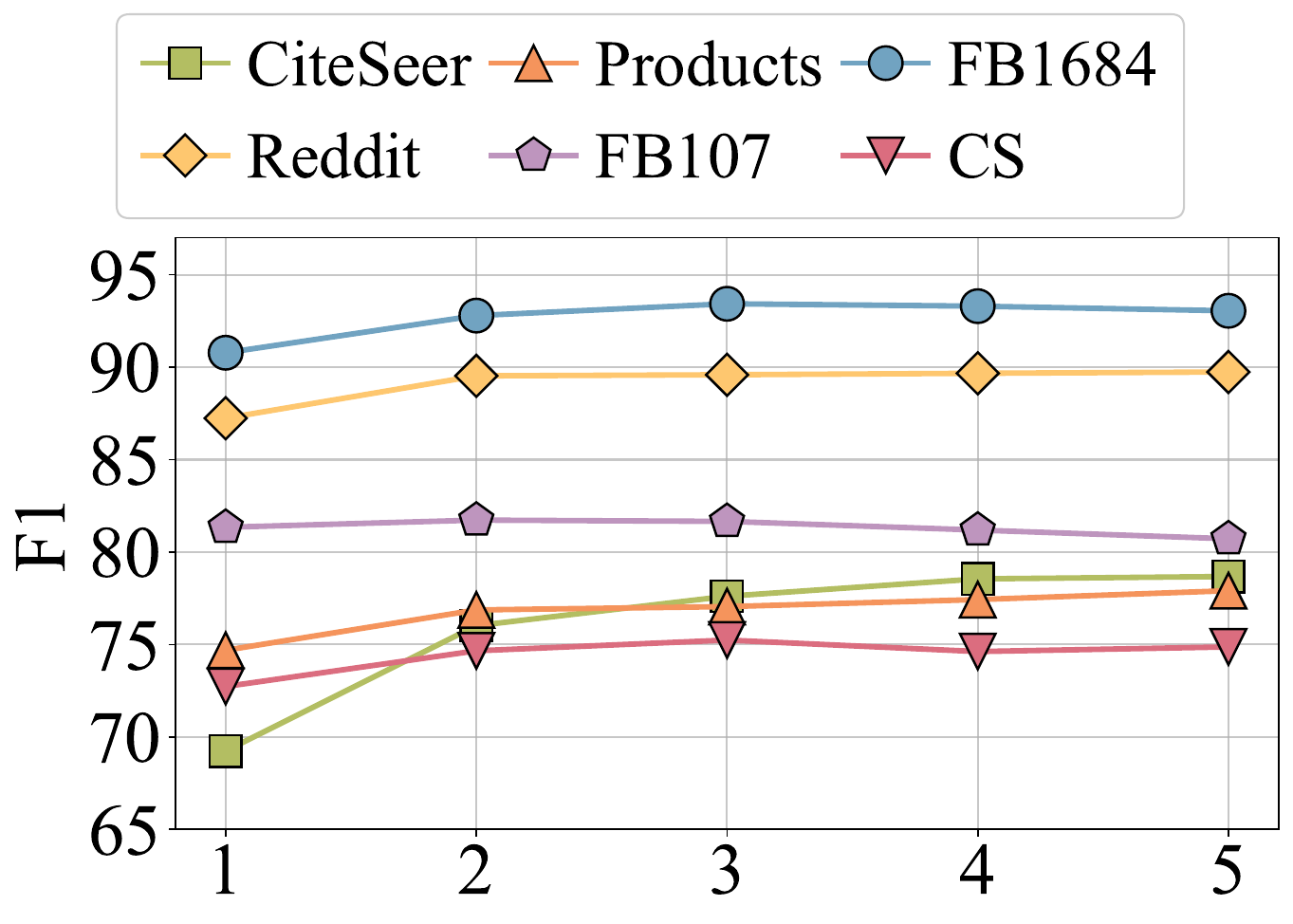}
  \vspace{-5.5mm}
  \caption{CS -- \# Graphs.}
  \vspace{1mm}
  \label{fig:n_graphs_cs}
\end{subfigure}

\par

% -------- Row 2 --------
\begin{subfigure}[b]{0.325\linewidth}
  \centering
  \includegraphics[width=\linewidth]{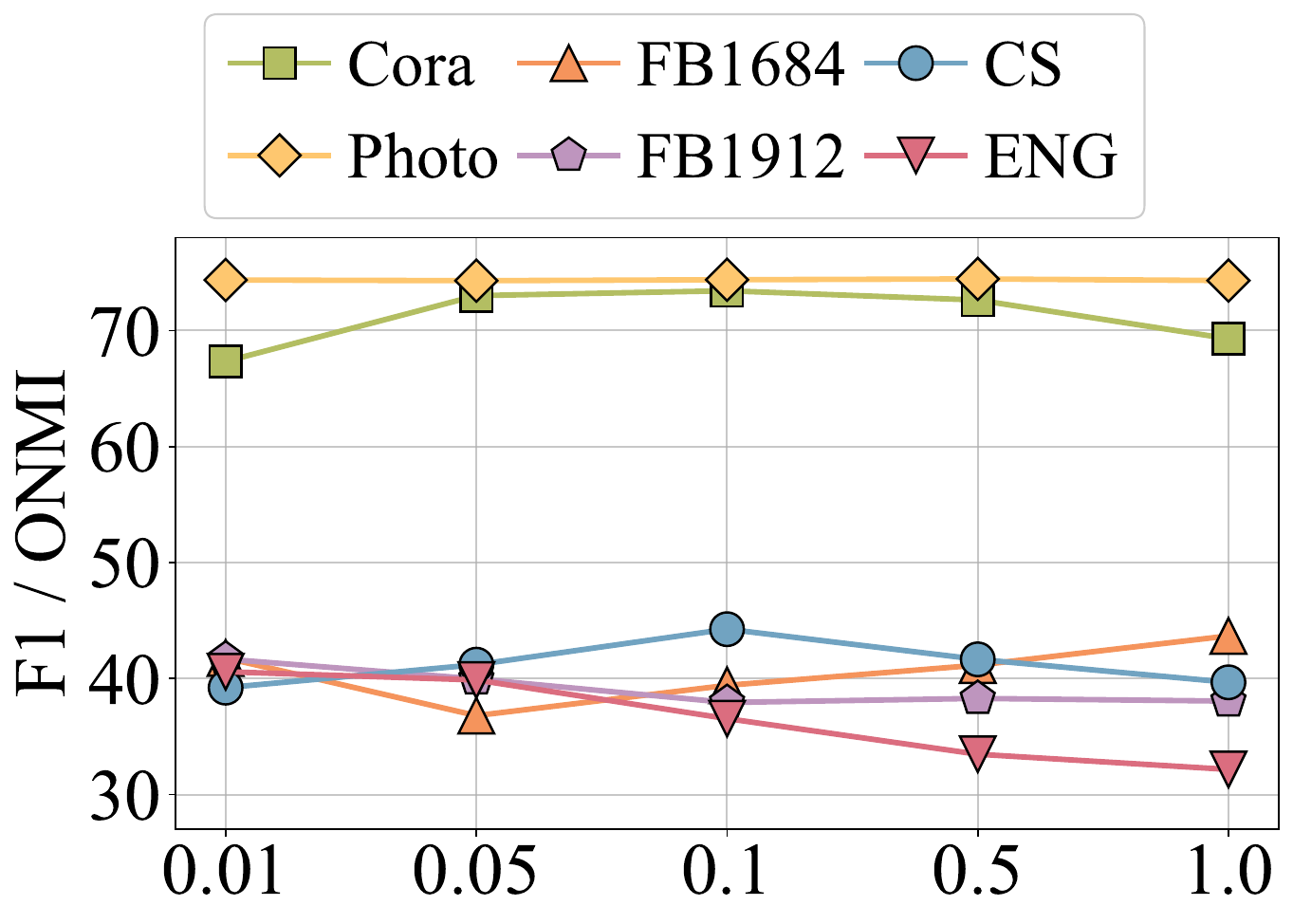}
  \vspace{-5.5mm}
  \caption{CD -- $\beta$.}
  \vspace{1mm}
  \label{fig:alpha_cd}
\end{subfigure}\hfill
\begin{subfigure}[b]{0.325\linewidth}
  \centering
  \includegraphics[width=\linewidth]{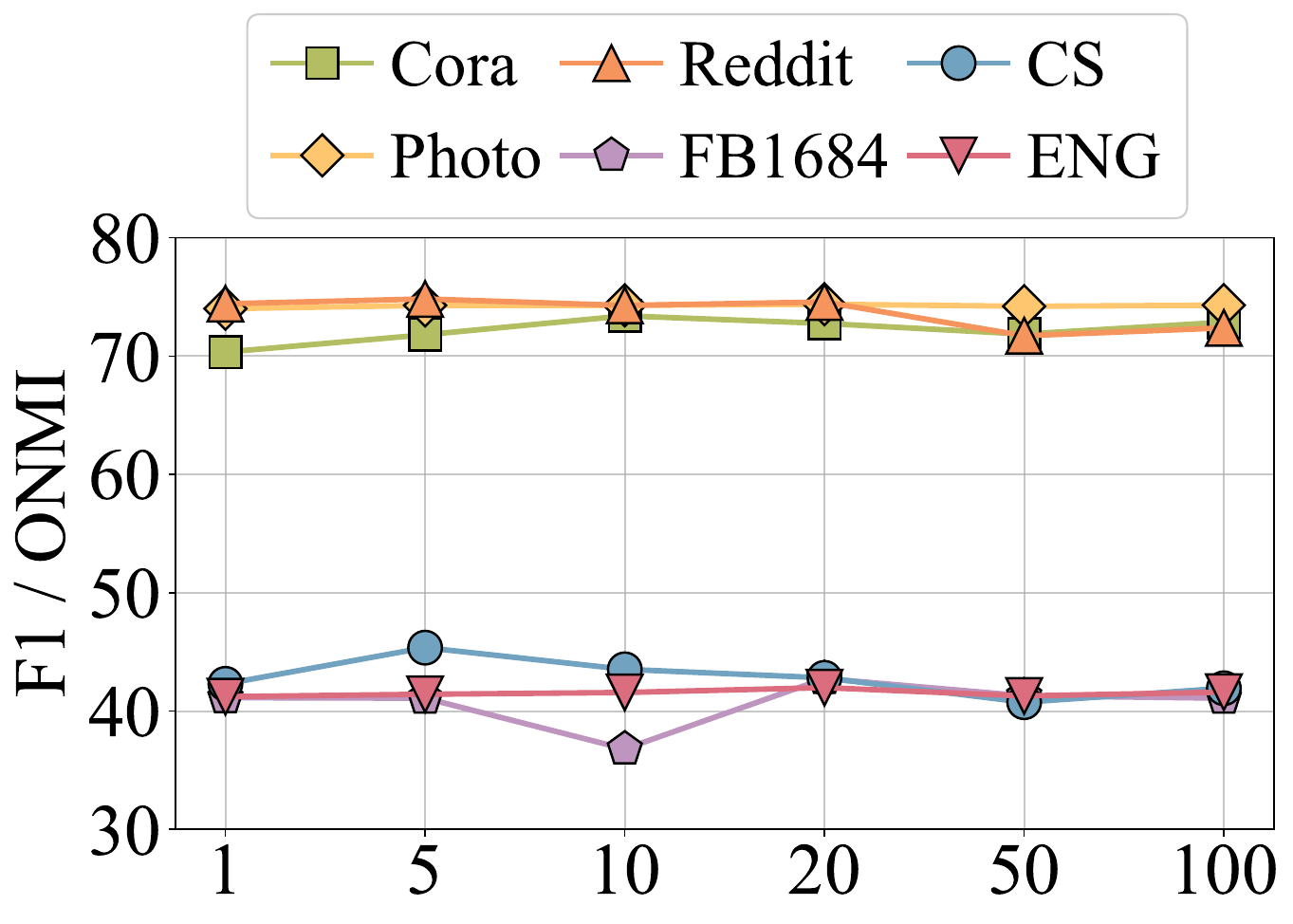}
  \vspace{-5.5mm}
  \caption{CD -- \# Prompts.}
  \vspace{1mm}
  \label{fig:n_prompt_cd}
\end{subfigure}\hfill
\begin{subfigure}[b]{0.325\linewidth}
  \centering
  \includegraphics[width=\linewidth]{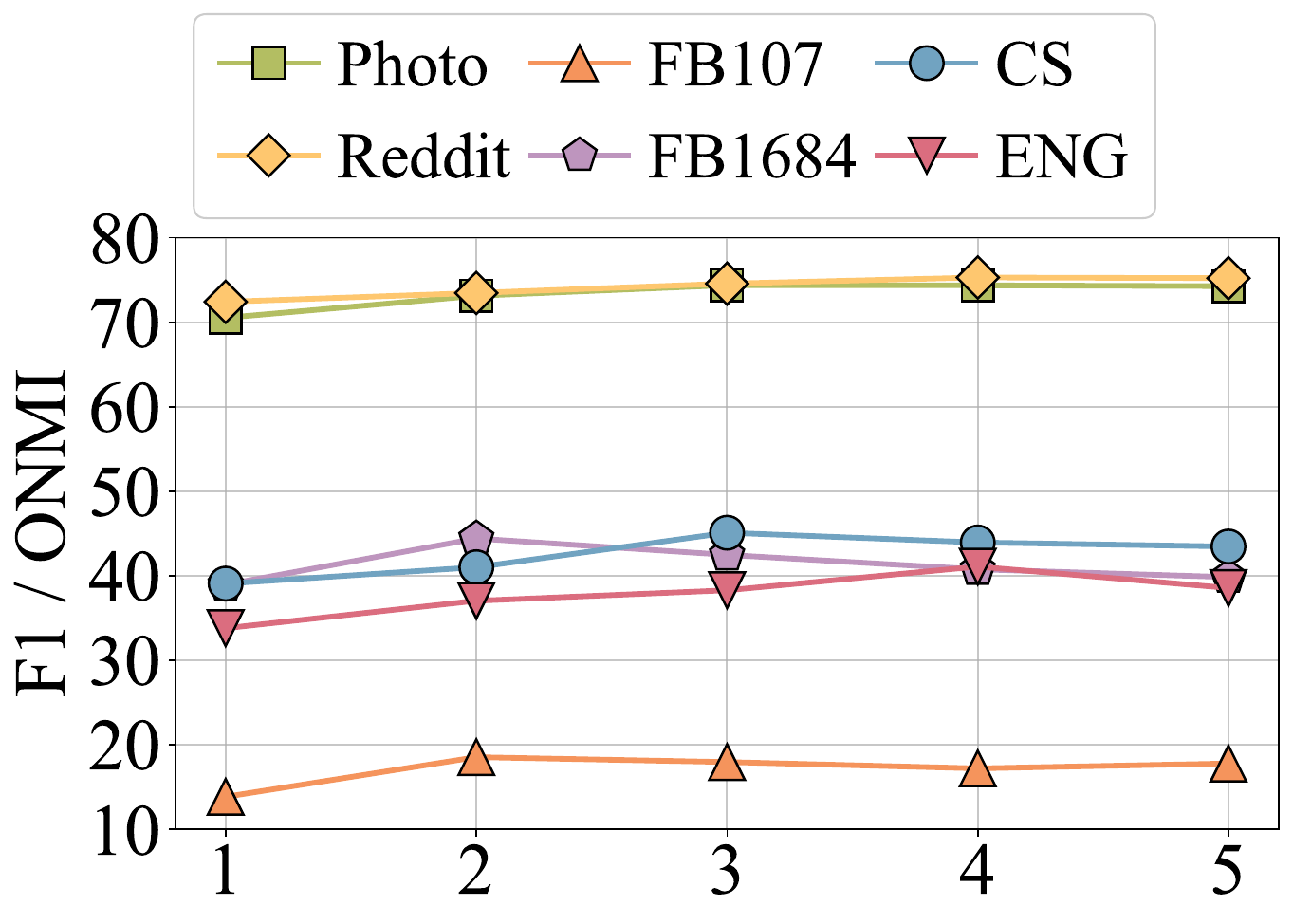}
  \vspace{-5.5mm}
  \caption{CD -- \# Graphs.}
  \vspace{1mm}
  \label{fig:n_graphs_cd}
\end{subfigure}

\vspace{-2mm}
\caption{Hyper-parameter Analysis.}
\vspace{-2mm}
\end{figure}

\noindent \textbf{Trade-off Parameter $\beta$.}
We investigate the impact of the trade-off parameter $\beta$ during the adaptation stage, and evaluate values of 0.01, 0.05, 0.1, 0.5, and 1, where $\beta$ balances the task loss and CMMD loss.
As illustrated in \autoref{fig:alpha_cs} and \autoref{fig:alpha_cd}, \model achieves the best performance when $\beta$ falls within a moderate range (0.05–0.5), while remaining stable and competitive across all evaluated values.

\noindent \textbf{Number of Basis Prompts.}
\autoref{fig:n_prompt_cs} and \autoref{fig:n_prompt_cd} present the performance of \model under different numbers of basis prompts (1, 5, 10, 20, 50, and 100).
\model remains strong and competitive across all settings, with the best results typically achieved with 5 to 50 prompts, indicating that simply increasing tunable parameters does not necessarily improve performance and that balancing model adaptability and parameter efficiency is more important.

\noindent \textbf{Number of Pre-training Graphs.}
To evaluate how the number of pre-training graphs affects performance, we progressively incorporate 1 to 5 pre-training datasets, as listed in \autoref{tab:datasets}.
As shown in \autoref{fig:n_graphs_cs} and \autoref{fig:n_graphs_cd}, performance gradually approaches a plateau as the number increases, beyond which it no longer improves.
This validates the data efficiency of \model, indicating that a limited number of pre-training datasets is sufficient for optimal performance.

\begin{table}[tb]
\setlength{\tabcolsep}{4.0pt}
\centering
\caption{Ablation Study.}
\vspace{0mm}
\renewcommand{\arraystretch}{0.90}
\begin{tabular}{c|cc|cc|cc}
\toprule
\textbf{Datasets} & \textbf{PubMed} & \textbf{CS} & \textbf{Cora} & \textbf{Photo} & \textbf{FB1684} & \textbf{FB1912} \\
Tasks & CS & CS & CD & CD & CD & CD \\
\midrule
w/o Pre-train   & 92.30 & 71.98 & 65.31 & 57.98 & 28.71 & 37.31 \\
w/o Fusion      & 92.86 & 73.32 & 72.53 & 71.66 & 40.36 & 40.30 \\
w/o Coh Prompt  & 93.44 & 72.77 & 55.23 & 72.74 & 40.38 & 40.22 \\
w/o Adp Prompt  & 91.81 & 72.86 & 37.59 & 59.97 & 31.96 & 40.75 \\
\midrule
Full Model      & \textbf{94.51} & \textbf{74.85} & \textbf{73.43} & \textbf{74.38} & \textbf{42.21} & \textbf{41.11} \\
\bottomrule
\end{tabular}
\vspace{-1mm}
\label{tab:ablation_studies}
\end{table}

\section{Conclusion}

In this paper, we address the long-standing gap between community search and detection by introducing \model, a unified framework for enhancing multi-domain knowledge transferability across community search and detection tasks on diverse datasets.
\model leverages a novel Domain-aware Specialization (DAS) stage, supported by Universal Graph Learning (UGL) stage, both of which incorporate community-aware augmentation to capture structural and semantic information at local and global levels.
UGL extracts knowledge from multiple domains with domain-specific models and community-aware objectives.
DAS transfers knowledge to target graphs through adaptation prompts and a lightweight projector.
\model demonstrates effectiveness and efficiency across multiple tasks, as validated by experiments on $16$ datasets and comparisons against 22 baselines.
This generalizability reveals its potential to serve as a foundation model for a wide range of subgraph-level tasks.

\section*{Acknowledgment}

% Hanchen Wang is supported by the Australian Research Council (ARC) under Grant DE250100226.
% Wenjie Zhang is supported by the Australian Research Council (ARC) under Grant DP260100689.
Hanchen Wang is supported by ARC DE250100226.
Wenjie Zhang is supported by ARC DP260100689.

\bibliographystyle{IEEEtran}
\bibliography{references}

\end{document}